# Electrical detection of plasmon-induced isomerization in molecule-nanoparticle network devices

Didier Stiévenard,[1] David Guérin,[1] Stéphane Lenfant,[1] Gaëtan Lévêque,[1] Christian A. Nijhuis[2,3,4] & Dominique Vuillaume[1]*.

1) Institute of Electronics, Microelectronics and Nanotechnology (IEMN),
CNRS, Université de Lille,
Avenue Poincaré, F-59652cedex, Villeneuve d'Ascq, France.

2) Department of Chemistry, National University of Singapore, 3 Science Drive 3, 117543 Singapore, Singapore.

3) Centre for Advanced 2D Materials and Graphene Research Centre, National University of Singapore, 6 Science Drive 2, 117546 Singapore, Singapore.

4) NUSNNI-Nanocore, National University of Singapore, Singapore 117411, Singapore.

* Corresponding author : dominique.vuillaume@iemn.fr

**We use a network of molecularly linked gold nanoparticles (NPSAN: nanoparticles self-assembled network) to demonstrate the electrical detection (conductance variation) of a plasmon-induced isomerization (PII) of azobenzene derivatives (azobenzene bithiophene : AzBT). We show that PII is more efficient in a 3D-like (cluster-NPSAN) than in a purely two-dimensional NPSAN (i.e., a monolayer of AzBT functionalized Au NPs). By comparison with usual optical (UV-visible light) isomerization of AzBT, the PII shows a faster (a factor > ~ 10) isomerization kinetics. Possible PII mechanisms are discussed: electric field-induced isomerization, two-phonon process, plasmon-induced resonant energy transfer (PIRET), the latter being the most likely.**

Networks of metal nanoparticles linked by molecules (or for short NPSAN: nanoparticles self-assembled network) are very valuable and useful in molecular electronics to study fundamental electron transport mechanisms, as well as to assess their potential applications in nanoelectronics.[1-2] For instance, NPSANs with simple molecules (alkyl chains, short π-conjugated oligomers) have been used to gain new insights in metal-insulator transitions,[3-4] plasmon enhanced photoconductance,[5-7] and co-tunneling,[8-11] while NPSANs with relative complex molecules give access to functional devices such as optically[12-15] and redox[16] driven molecular switches, molecular memory and negative differential resistance,[17-18] and unconventional computing approaches. Bose et al. demonstrated the training of molecule/ nanoparticle networks via genetic algorithms in an alkylthiol capped Au NPs network dominated by Coulomb blockade below 5K.[19] Viero et al. demonstrated that NPSANs of Au NPs functionalized by optically driven molecular switches exhibit both optically driven reconfigurable logic operations and reconfigurable strongly non-linear electron transport and dynamic behaviors (high-harmonic generation) required for reservoir computing.[20]

In molecular electronic plasmonics (see Ref. [21] for a review), molecules chemically grafted, or deposited as thin films, on metallic nanostructures are used to modify the local surface plasmon (LSP) frequency depending on their molecular states (i.e., redox state, conformation or configuration, dipole moment changing the dielectric constant around the metal nanostructures), which, in turn, can be controlled via an applied bias.[22-29] Optical inputs can also be used, for example, the photoisomerization of photochromic molecules (azobenzene, diarylethene, or spiropyran derivatives) directly bound to metallic nanoparticles or nanostructures changed the frequency of the LSP resonance.[22, 26-28] This plasmonic detection of the isomer state of the photochrome molecules has been observed at a single nanoparticle level.[30] Plasmon resonances can be also detected by electric means with tunnel junctions via optical rectification where the surface plasmon modes modulate the tunneling current flowing across the junction.[31-32] Several authors reported an increase of the tunnel current measured through various metal/molecules/metal junctions when these devices are irradiated by light with the same frequency as the plasmon resonance frequency;[32-37,7, 38] this phenomenon (also called plasmon-assisted tunneling (PAT)) can be modeled based on a photon-assisted tunneling model initially introduced by Tien and Gordon.[39] Plasmonic electronics is also interesting for applications in quantum plasmonics.[40-41] For example, the conductivity of an organic semiconducting thin film is increased when strongly coupled to the plasmonic modes of a nanostructured metallic layer,[42] and tunneling charge transfer plasmon (tCTP), resulting from quantum interactions between tunneling electrons and plasmonic nanostructures, was recently demonstrated in molecular tunnel junctions of various molecules connected between two NPs.[43-44]

Albeit the isomerization of molecules attached onto a NPs induced by LSP excitation (here referred to as PII: plasmon-induced isomerization) has been observed from purely optical methods (Raman or fluorescence spectroscopies),[45-48] the modulation of the electrical conductance of a molecularly functionalized NP network induced by PII of the molecules decorating the NPs has not been reported. Here, we combine NPSANs and molecular plasmonics to demonstrate the electrical detection (conductance variation) of PII of azobenzene derivatives (azobenzene bithiophene: AzBT). These molecules are linked to gold nanoparticles and embedded in a planar nanodevice consisting of NPSANs contacted by nanoelectrodes separated by 100-200 nm inter-electrode gaps. We show that the PII effect is observed in the cluster-NPSANs due to an efficient enhancement of the LSP resonance through an antenna effect, while a weak effect is observed in 2D-NPSANs (i.e., a monolayer of AzBT functionalized Au NPs). The usual optical isomerization of AzBT (UV-visible light) is also electrically detected in both 2D- and cluster-NPSANs with similar kinetics for both types of devices. In sharp contrast, due to PII, the switching kinetics are faster by a factor of ~10 .

In this study, we focus on AzBT (Fig. 1-a)-capped gold nanoparticles (AuNPs, Fig. 1-b) that were synthesized by ligand exchange reactions from a suspension of citrate-capped AuNPs, 10 nm in diameter, in the presence of AzBT. The details of the synthesis of AzBT molecules, the preparation of the AzBT NPSANs, and the deposition of the NPSANs on two nanoelectrodes, have been reported by us before[15, 49-50] and are briefly described in the Methods. The choice of this molecule is motivated by the fact that high conductance ratios between the two photoisomerizable states (*trans* and *cis*) have been observed when these molecules are self-assembled as a monolayer on Au surfaces[50] and embedded in NPSANs.[15] Since the plasmonic effect is very sensitive to the geometry of the NP assemblies, which act as antennas for the detection and the amplification of the incident electromagnetic waves,[51-52] we fabricated two types of devices: monolayer and multilayer NPSANs were deposited within the electrodes leading to the hereafter named 2D-NPSAN (Fig. 1-c) and cluster-NPSAN (Fig. 1-d).

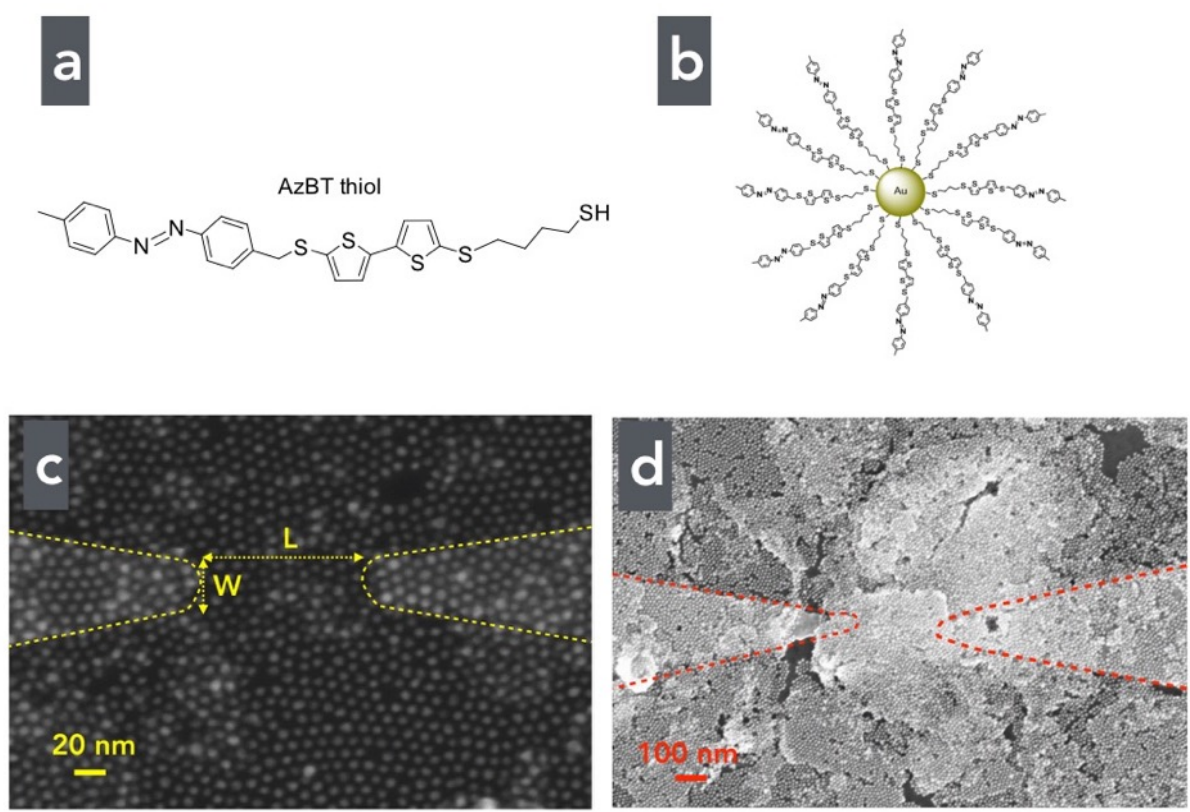

*Figure 1. (a) Molecular structure of the azobenzene-dithiophene-alkylthiol (AzBT). (b) Schematic illustration of Au NPs functionalized with AzBT. (c) Scanning electron microscope images of 2D-NPSAN spanning the L ∼150 nm gap between two electrodes (W ∼40 nm) and (d) a cluster-NPSAN nanodevice with ∼ 3-5 layers of AzBT-NPs (inter-electrode distance of ∼*

*200 nm). Dashed lines (guide to the eyes) indicate the location of the electrodes underneath the NPSANs*

## Effect of the plasmonic excitation.

To study the effect of a plasmonic excitation, we preconfigured the NPSANs with the AzBT such that a maximum number of molecules were in their *trans* or *cis* state by illuminating the NPSANs at 365 nm and 470 nm (for 1-1.5 h), respectively. This pre-configuration is followed by a period in dark (typically 10-20 min) until a stable current was reached, then we illuminated devices at 590 nm to excite the LSP for 3-15 min until a stable current was observed (see Methods). We recorded the time evolution of the current through the NPSANs at a constant dc voltage of 2.5 V (see Methods). This voltage was chosen because the junctions remain stable and the currents albeit weak are above the noise levels.

For the 3D NPSANs preconfigured with the AzBT *trans* state (illumination at 470 nm for ca. 1h), Fig. 2-a shows the current vs. time, $I(t)$, curves recorded under LSP excitation starting from devices in the dark followed by illumination at 590 nm and, subsequently, turning the light source off rendering the devices again in the dark. Fig. 2-b shows the results for the NPSANs preconfigured with AzBT in the *cis* isomer (illumination at 365 nm for ca. 1h). The current increases when the AzBT molecules are preconfigured in their *trans* state (Fig. 2-a) while it decreases for the *cis* state (Fig. 2-b). The time constants are $90 \pm 9$ s and $103 \pm 10$ s (i.e., not significantly different) for both types of devices, respectively. The value of the difference in current before and after illumination, $\Delta I$, for the *trans*-AzBT devices is 4 pA, but the value of $\Delta I$ of -70 pA is more than one order of magnitude larger for the *cis*-AzBT devices. Here, the relative current amplitude variations are 10 to 20 % of the steady-state currents before illumination. Moreover, when the 590 nm light source is switched off, the current remains stable.

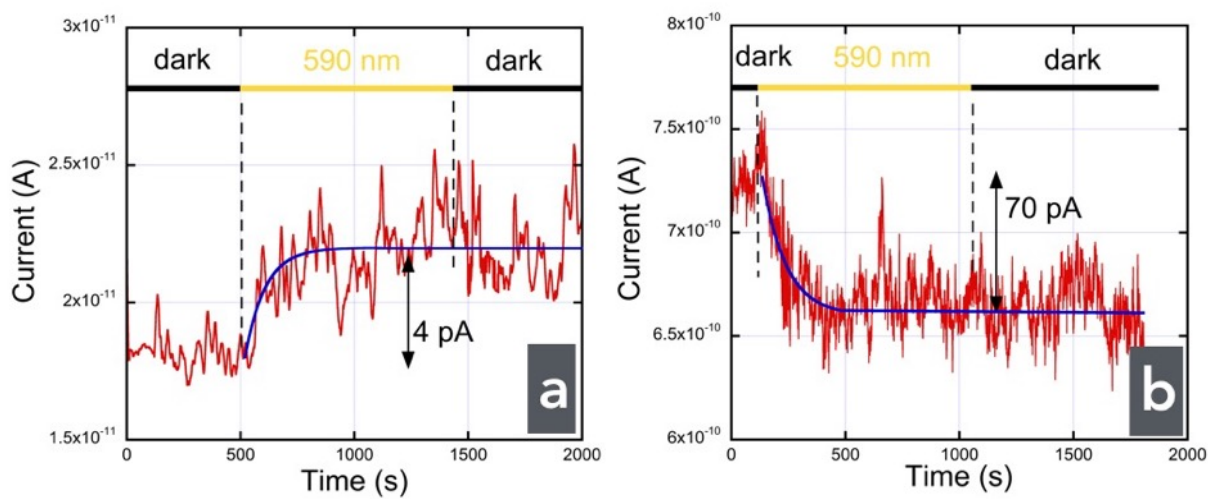

*Figure 2. Measured current vs. time traces under 590 nm excitation for cluster-NPSAN with the AzBT in (**a**) the trans and (**b**) the cis configurations measured at 2.5 V with an electrode spacing of ~200 nm (see Fig. 1d). Blue lines are fits to single exponential functions (time constants of 90 ± 9 s and 103 ± 10 s, respectively). The vertical arrows indicate the change in the current ΔI.*

Drifts in current (during the illumination and dark periods), which are larger than the current variations induced by the LSP excitation, complicate the same experiments in the 2D-NPSANs (see Figs. S1-S2 in the supporting information). Consequently, figure 3-a shows the corrected current adjusted for this drift (see supporting information) recorded from the 2D NPSANs with the AzBT molecules preconfigured in their *trans* state. It was not possible to do the same experiment when the AzBT molecules are preconfigured in their *cis* state because the current immediately decreased when the UV light was switched off (Fig. S2, supporting information). We have not obtained a stable dark current (as in cluster-NPSANs, Fig. 2) before turning on the LSP excitation light at 590 nm. This is likely due to the metastable character of the *cis* state at room temperature in the dark, which spontaneously returns to the *trans* state.[53] We rationalize the feature that a more stable *cis* state in the dark is observed for the cluster-NPSANs by rising the hypothesis that the NPs are more closely packed in the cluster than in the 2D layer, reducing the free space around the AzBT molecules and consequently reducing the *cis-to-trans* spontaneous back isomerization efficiency. For example, it was observed that the *cis* state, in dark at room temperature, can be stable up to several hours in densely packed azobenzene SAMs.[50] Under this condition, when we apply the LSP excitation, a large fraction of the AzBT molecules have turned back to the *trans* state and we again observe a current increase (Fig. 3-b) as in the case with the preconfigured *trans* state (Fig. 3-a). Exponential fits give rising time constants of $22 \pm 4$ s and $63 \pm 7$ s in both cases, respectively. The value of $\Delta I$ is in the range 1 to 4 pA, i.e., relative current variations around 0.5 % of the steady-state currents, much lower than in the case of the cluster-NPSANs.

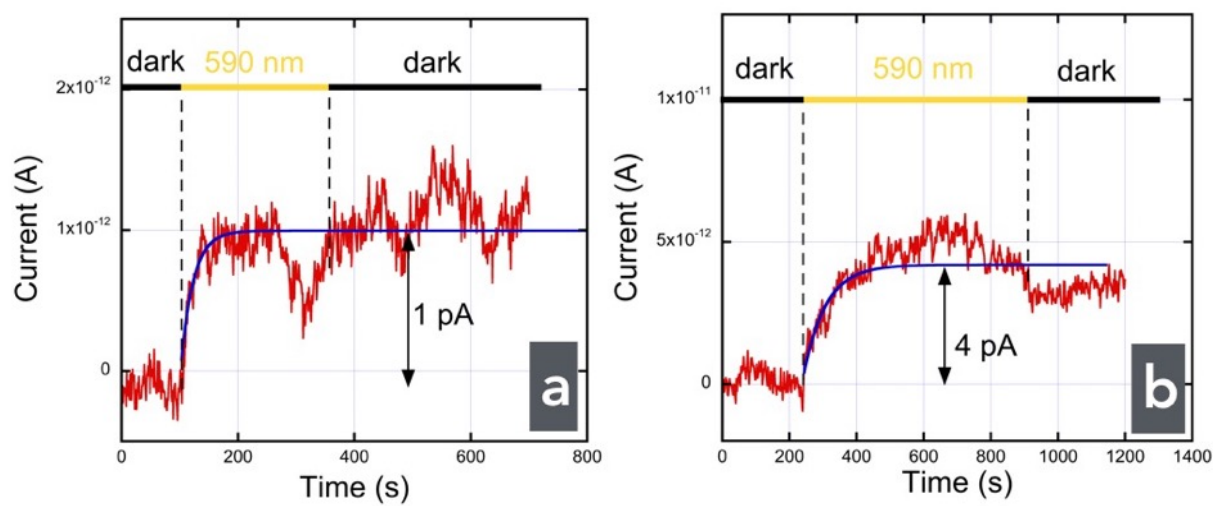

*Figure 3. Current vs. time traces (corrected from drift) under 590 nm excitation for 2D-NPSAN with the AzBT in (**a**) the trans and (**b**) the unstable cis configurations measured 2.5 V*

*with an electrode spacing of 150 nm (Fig. 1-c). Blue lines are fits to single exponential functions (time constants 22 ± 4 s and 63 ± 7 s, respectively). The vertical arrows show the change in the current ΔI.*

**UV and visible light photoisomerizations.**

For the sake of comparison, we also measured the kinetics of the AzBT photoisomerisation in the NPSANs under usual UV and visible light illuminations[15] and we examined the respective behaviors of the 2D-NPSANs and cluster-NPSANs. Figures 4-a and b give typical *I*(t) curves at 2.5 V for (a) the *trans* to *cis* (365 nm light excitation) and (b) the *cis* to *trans* (470 nm light excitation) AzBT photoisomerization in the 2D-NPSANs using preconfigured devices (Methods). Figures 4-c and d show that a similar behavior is observed for the cluster-NSANs but with higher currents (as expected due to a more important number of current pathways). However, in all these cases, we observed multi-exponential kinetics with characteristic time constants ranging from few tens to few $10^4$ s (see details and values in the supporting information, Table S1). The main observation is that these kinetics are markedly different from those measured under excitation at 590 nm (Figs. 2 and 3), which indicates that different mechanisms are involved.

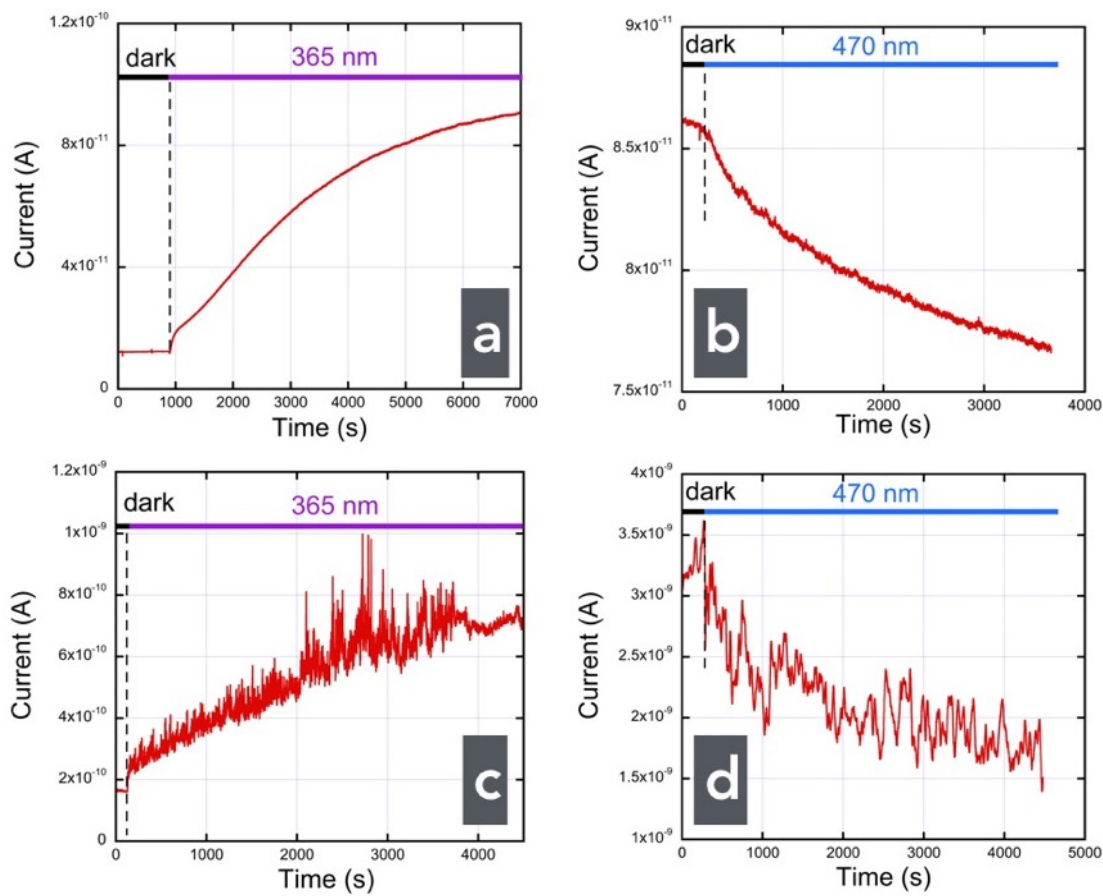

*Figure 4. Current vs. time traces recorded from 2D-NPSANs, (a) for the trans to cis isomerization (under 365 nm illumination), and (b) for the cis to trans isomerization (under 470 nm illumination). Current vs. time traces recorded from cluster-NPSANs, (c) for the trans to cis isomerization (under 365 nm illumination ), (d) for the cis to trans isomerization (under 470 nm illumination at 2.5 V electrode spacing of ~150 nm and ~ 200 nm, respectively).*

**Discussion of mechanisms under plasmonic excitation.**

Several hypotheses have to be considered to explain the current increase under light (590 nm) illumination. Since there is no optical absorbance of the AzBT molecule at 590 nm,[49] a direct *trans-to-cis* and *cis-to-trans* photoisomerization can be ruled out for the observed changes in current under illumination shown in Figs. 2 and 3. Since the optical excitation energy at 590 nm (2.1 eV) is lower than the molecule HOMO-LUMO gap (3.4 eV[49]), we do not believe that photocurrents, as reported for networks of NPs functionalized by conjugated molecules[54] and molecular junctions,[55] are important in our devices. We also dismiss thermal effects associated with the absorption of the light as it has been reported for NP arrays involving the LSP in Ref. 6. In this latter case, an increase of the current of 5 pA was obtained under 4.1 kW/cm$^2$ illumination.[6] Considering that we use much lower power of 10 mW/cm$^2$, we can safely dismiss thermal effects. For these reasons, we believe that excitation of the molecules is possible *via* the LSP resonance of the gold NPs.

It is well known that plasmons can increase tunneling currents flowing across tunneling junctions; this process is called plasmon-assisted tunneling (PAT).[32-37] However, the slow rising response times (~ 20-100 s, Figs. 2 and 3) cannot be explained with PAT as tunneling events occur on 1fs time scales, plasmon lifetimes are on the order of tens of fs time scales. Consequently, when the light is switched off, the current must decrease almost immediately, which was not observed in our case. Moreover, we can estimate the PAT current using the same model as Vadai *et al.*[37], based on Tien-Gordon model.[39] The plasmon field is treated as a local potential $V_\omega$ across the NP-molecules-NP junctions, which oscillates at the plasmon frequency $\omega$. This potential is given by :[37]

$$V_\omega = \frac{2\hbar\omega}{e} \sqrt{\frac{G_{dc}(\omega) - G_{dc}(0)}{G_0 T(E_F \pm \hbar\omega)}} \qquad (1)$$

where $G_{dc}$ is the single molecule conductance with ($G_{dc}(\omega)$) and without ($G_{dc}(0)$) the light excitation, $G_0$ is the quantum of conductance (7.75 x 10$^{-5}$ S), $\hbar$ the reduced Planck constant, e the electron charge and $T(E_F \pm \hbar\omega)$ is the electron transmission probability through the junction for electrons at $E_F \pm \hbar\omega$. We calculated $V_\omega$, based on a classical electromagnetic model (see the supporting information) and estimated a typical value of 1.3x10$^{-7}$ V, leading to a PAT current to be around 10$^{-17}$ A at maximum (see supporting information) in the NPSANs (below the detection limit of our equipment, about 10 fA).

We propose that the observed variation of the current is due to a plasmon-induced isomerization (PII) of the AzBT molecules inserted in the NPSANs. This PII effect is observed for cluster-NPSANs devices with an enhanced effect compared to 2D NPSANS due to an antenna effect. We discuss several hypotheses to explain the AzBT isomerization by a LSP excitation.

(i) It was shown that the isomerization of azobenzene molecules can be triggered by an electric field instead of UV and visible light.[56] However, this mechanism requires electric fields in the range of few V/nm. Here the plasmonic field $E_\omega$ is around $3\times10^{-8}$ V/nm (see above, $V_\omega \sim 1.3\times10^{-7}$ V divided by an average gap of 4.5 nm between NPs[15]), well below the threshold required for an electric field-induced isomerization. Even if we assume a collective 3D antenna effect enhancing the local electric field by ca. 10 (the precise factor depending on the size of NP clusters and its organization, i.e. number of neighboring NPs), the field generated by the LSP excitation remains below this threshold. Moreover, the experimental conditions in Ref. [56] are very different from our case involving single molecule measured by STM in UHV (see details in the supplementary information).

(ii) A second hypothesis is a two-photon mechanism. This mechanism was observed by fluorescence experiments on Ag nanoparticles decorated with benzooxazine ring derivatives and diarylethene.[46-48] Here, the LSP excitation at 590 nm requires a significant molecular absorption at around 295 nm. For the AzBT molecules, the absorption at this wavelength is lower (about 1/3) than at the peak of maximum absorption (at 336 nm) but not zero.[49] However, at low light power excitation as in the present case, the probability of a two-photon absorption is several orders of magnitude weaker than direct absorption.

(iii) Plasmon-induced resonance energy transfer (PIRET)[57] is another possible mechanism that enables energy transfer from the LSP resonance towards the shorter wavelength of the isomerization bands of AzBT. The required spectral overlap between the LSP band and the absorbance of the AzBT molecules exists around 400-500 nm in the AzBT-NPSANs (see Fig. S7 in the supporting information). Distinguishing PIRET and 2-photon process would require measurements varying the light incident power (PIRET scale linearly with light power, albeit 2-photon process scale quadratically), which are not possible with our apparatus. The two-photon absorption probability is $10^{-4}$-$10^{-5}$ smaller than the one predicted to PIRET,[57] thus we conclude that PIRET is the most likely mechanism to explain the PII in AzBT-NPSANs. A plasmon energy transfer process from the NPs to the molecules was also proposed, albeit without molecular isomerization, to explain a light-increase conductance of molecular junctions of hemicyanime dyes.[7, 58] Nevertheless, these hypotheses deserve more detailed theoretical calculations, especially to understand the difference between 2D- and cluster-NPSANs.

The time constants measured for the PII (typically from 20 to 100 s, see Figs. 2 and 3) are lower compared to the UV-visible photoisomerization time constants (> $10^3$ s, Fig. 4 and Table S1 in the supplementary information, if we except the fast one in the case of the *trans* to *cis* isomerization). The light powers are not rigorously the same at the various wavelengths (see Methods) but cannot explain such large differences. Following theoretical calculations by Hush et al.,[59] and experimentally reported by Mahmoud,[45] the photo-isomerization time constant decreases with increasing the local dynamic electric field induced by the LSP. The observed decrease of PII time constants compared to usual UV-visible isomerization is qualitatively consistent with the dynamic increase of the local electric field induced by the LSP excitation. We also note that PII time constants observed in our work (20-100 s with a light power of ca. 10 mW, see Methods) are qualitatively consistent with the ones reported in Ref. [45] (ca. 8 minutes but at a lower light power of 1 mW, thus at a lower plasmonic electric field). Since the plasmonic electric field depends on the inter-NP distance, a related question is whether or not this distance changes upon AzBT isomerization. It is known that *trans-cis* isomerization of azobenzene functionalized NP embedded in a polymer matrix can induce a large and reversible change of the NP aggregation, due to large change of the dipole moments.[60] Atomic force microscopy images (supporting information, Fig. S9) of the AzBT-NPSANs with molecules in the *trans* and *cis* states show no measurable modification of the NP network morphology in our case of NPs deposited on a solid surface. Similarly, the modification of AzBT dipole upon *trans-cis* isomerization, interacting with the plasmonic dipole, can induce some change in the NP-molecule energy transfer. However, we have not seen any significant changes in the PII kinetics (Fig. 2 and Table 1 in the supplementary information). This is due to the fact that in our experiments, we detect the molecule isomerization kinetics, which is induced by the energy transfer, this latter being much faster (ps regime, see Ref. [57]) than the AzBT isomerization process itself.

## Conclusion

In summary, we demonstrate the electrical detection (change in conductance) of the plasmon-induced isomerization (PII) in a network of gold nanoparticles functionalized with azobenzene derivatives. We rule out several mechanisms, which can also induce conductance changes in molecular devices such as excitonic photocurrent, thermal photoconductance, and plasmon-assisted tunneling (PAT). We discuss several mechanisms for this plasmon-induced isomerization, among them plasmon-induced resonance energy transfer (PIRET) seems more likely than electric field-induced isomerization and two-phonon absorption. The plasmon-induced isomerization is faster (at least a factor 10) than the usual *trans-cis* isomerization triggered by UV-visible light, which may have practical outcomes for light-driven molecular memories and light reconfigurable molecular circuits.

**Methods.**

The coplanar titanium/gold (1nm/10nm thick) electrodes connecting the NPSANs were fabricated using standard electron-beam lithography and electron-beam metal deposition on silicon substrate covered by a 220 nm thick thermally grown silicon dioxide.[15] The synthesis of AzBT molecules and AzBT functionalized gold NPs (10 nm in diameter) were performed following protocols already used in our previous works.[15, 49] The 2D-NPSANs were prepared according to already reported methods.[61-62] More precisely here, we followed the method of Santhanam et al.,[62] by forming a Langmuir film of AzBT-AuNPs floating on the water surface and transferring it by dip-coating onto the lithographed substrates with the electrodes.[15] During the formation of the Langmuir film by evaporation of a solution of AzBT-AuNPs in TCE (1,1,2,2-tetrachloroethane), the addition of an excess of hexane to the solvent at the water surface produced the aggregation and formation of multilayers of AzBT-AuNPs, which are transferred on the electrode to fabricate the cluster-NPSANs. Detailed characterizations of the NPSANs were reported in our previous work.[15] These included: AFM images to determine the average NP diameter (10.2 nm), inter-NP distances (with AzBT in the *trans* state, 4.5 nm), UV-visible absorption spectra of the AzBT-NPs in solutions and the AzBT-NPSANs on a substrate (*trans* and *cis* states) to check the reversible isomerization of the AzBT molecules, XPS analysis to check the chemical composition of the AzBT-NPs from which we deduced the density of ca. 3.2 molecules/nm$^2$ (i.e. about 2560 AzBT molecules per NP).

The electrical measurements were performed with an Agilent4156C semiconductor parameter analyzer. To avoid any degradation of the organic molecules during the measurements, the nanoelectrodes were contacted with a probe station (Suss Microtec PM-5) inside a glove box (MBRAUN) under a atmosphere of nitrogen (less than 0.1 ppm of oxygen or water vapor). Before light illumination, current voltage (I-V) curves were recorded in dark to check the quality of the measured devices. Then, current-time I(t) were recorded, under a constant bias voltage (2.5 V) and under a specific sequence of light excitation (see below). For the light exposures, an optical fiber, with a 400μm diameter, was brought near the samples inside the glovebox. We used three power LED from Thorslabs, with the following characteristics: i) 365 nm, bandwidth of 10 nm, for the *trans*-to-*cis* isomerization, ii) 470 nm, bandwidth of 20 nm, for the cis-to-trans isomerization and iii) 590 nm, bandwidth of 80 nm for the gold plasmonic excitation. The light power values were measured with a calibrated Optical Power and Energy meter PM200 (Thorlabs). They were 10.2 mW, 8.6 mW, and 11.5 mW, on the devices at the output of the optical fiber for the 365, 470 and 590 nm sources, respectively.

After the synthesis of the molecules, the AzBT molecules are mainly in the *trans* isomer.[63] Before we started our experiments, we ensured all the devices were preconfigured in the either the *cis* or *trans* state as follows. Pre-configuration in *trans (cis,* respectively): the devices were illuminated with the 470 nm  (365 nm, respectively) source for typically 1-1.5h. Then we let the devices in the dark for 10-20 min, before we switched on the 595 nm source (LSP excitation) for 3-15 min till steady state currents were observed.

## Associated content.

Supplementary information is available in the online version of the paper. The supplementary information includes: Drift base line corrections, kinetics of the UV-vis photoisomerizations, PAT current calculations, UV-visible absorption spectrum, plasmonic simulations, AFM images of the NPSANs.

Correspondence and request for materials should be addressed to D.V.

## Author information.

*Correspinding author*

dominique.vuillaume@iemn.fr

*ORCID*

D. Vuillaume: 0000-0002-3362-1669
D. Stievenard: 0000-0003-2748-0254
C. Nijhuis: 0000-0003-3435-4600

S. Lenfant: 0000-0002-6857-8752

G. Lévêque: 0000-0003-1626-8207

*Author contributions*

S.L. and D.G. fabricated the devices, D.G. synthesized the AzBT-AuNPs and NPSANs. D.S. performed all the electro-optical measurements, and analysed the data with D.V. S.L. performed the SEM measurements. G.L. performed the simulations. C.A.N. discussed the results and the manuscript. D.S. and D.V. wrote the paper, all the authors discussed the results and contributed to the preparation of the manuscript and revision toward its final form.

*Notes*

The authors declare no competing financial interests.

## Acknowledgements.

The French nanotechnology network RENATECH supported the nanotechnology work in clean room. We thank Philippe Blanchard (MOLTECH-Anjou, CNRS, Univ. Angers) for the synthesis of AzBT molecules. CAN acknowledges the Ministry of Education (MOE) of Singapore for supporting this research under award No. MOE2015-T2-2-134. We thank Y. Viero for the AFM images.

## References

1. Liao, J.; Blok, S.; van der Molen, S. J.; Diefenbach, S.; Holleitner, A. W.; Schönenberger, C.; Vladyka, A.; Calame, M., Ordered nanoparticle arrays interconnected by molecular linkers: electronic and optoelectronic properties. *Chem. Soc. Rev.* **2015,** *44*, 999-1014.
2. Zabet-Khosousi, A.; Dhirani, A.-A., Charge Transport in Nanoparticle Assemblies. *Chemical Reviews* **2008,** *108* (10), 4072-4124.
3. Zabet-Khosousi, A.; Trudeau, P.-E.; Suganuma, Y.; Dhirani, A.-A.; Statt, B., Metal to Insulator Transition in Films of Molecularly Linked Gold Nanoparticles. *Phys. Rev. Lett.* **2006,** *96*, 156403.
4. Tie, M.; Dhirani, A. A., Conductance of molecularly linked gold nanoparticle films across an insulator-to-metal transition: From hopping to strong Coulomb electron-electron interactions and correlations. *Phys. Rev. B* **2015,** *91*, 155131.
5. Bernard, L.; Kamdzhilov, Y.; Calame, M.; van der Molen, S. J.; Liao, J.; Schönenberger, C., Spectroscopy of Molecular Junction Networks Obtained by Place Exchange in 2D Nanoparticle Arrays. *J. Phys. Chem. C* **2007,** *111* (50), 18445-18450.
6. Mangold, M. A.; Weiss, C.; Calame, M.; Holleitner, A. W., Surface plasmon enhanced photoconductance of gold nanoparticle arrays with incorporated alkane linkers. *Appl. Phys. Lett.* **2009,** *94*, 161104.

7. Pourhossein, P.; Vijayaraghavan, R. K.; Meskers, S. C. J.; Chiechi, R. C., Optical modulation of nano-gap tunnelling junctions comprising self-assembled monolayers of hemicyanine dyes. *Nat Commun* **2018,** *7* 11749.
8. Tran, T.; Beloborodov, I.; Lin, X.; Bigioni, T.; Vinokur, V.; Jaeger, H., Multiple Cotunneling in Large Quantum Dot Arrays. *Phys Rev Lett* **2005,** *95* (7), 076806.
9. Pauly, M.; Dayen, J. F.; Golubev, D.; Beaufrand, J.-B.; Pichon, B. P.; Doudin, B.; Bégin-Colin, S., Co-tunneling Enhancement of the Electrical Response of Nanoparticle Networks. *Small* **2012,** *8*, 108-115.
10. Dayen, J. F.; Devid, E.; Kamalakar, M. V.; Golubev, D.; Guédon, C.; Faramarzi, V.; Doudin, B.; van der Molen, S. J., Enhancing the Molecular Signature in Molecule-Nanoparticle Networks Via Inelastic Cotunneling. *Adv. Mater.* **2013,** *25*, 400-404.
11. Tran, T. B.; Beloborodov, I. S.; Hu, J.; Lin, X. M.; Rosenbaum, T. F.; Jaeger, H. M., Sequential tunneling and inelastic cotunneling in nanoparticle arrays. *Phys. Rev. B* **2008,** *78*, 075437.
12. Liao, J.; Bernard, L.; Langer, M.; Schönenberger, C.; Calame, M., Reversible Formation of Molecular Junctions in 2D Nanoparticle Arrays. *Adv. Mater.* **2006,** *18* (18), 2444-2447.
13. Matsuda, K.; Yamaguchi, H.; Sakano, T.; Ikeda, M.; Tanifuji, N.; Irie, M., Conductance Photoswitching of Diarylethene−Gold Nanoparticle Network Induced by Photochromic Reaction. *The Journal of Physical Chemistry C* **2008,** *112* (43), 17005-17010.
14. van der Molen, S. J.; Liao, J.; Kudernac, T.; Agustsson, J. S.; Bernard, L.; Calame, M.; van Wees, B. J.; Feringa, B. L.; Schönenberger, C., Light-controlled conductance switching of ordered metal-molecule-metal devices. *Nano Letters* **2009,** *9* (1), 76-80.
15. Viero, Y.; Copie, G.; Guerin, D.; Krzeminski, C.; Vuillaume, D.; Lenfant, S.; Cleri, F., High Conductance Ratio in Molecular Optical Switching of Functionalized Nanoparticle Self-Assembled Nanodevices. *J. Phys. Chem. C* **2015,** *119*, 21173-21183.
16. Liao, J.; Agustsson, J. S.; Wu, S.; Schönenberger, C.; Calame, M.; Leroux, Y.; Mayor, M.; Jeannin, O.; Ran, Y.-F.; Liu, S.-X.; Decurtins, S., Cyclic Conductance Switching in Networks of Redox-Active Molecular Junctions. *Nano Lett* **2010,** *10* (3), 759-764.
17. Mangold, M. A.; Calame, M.; Mayor, M.; Holleitner, A. W., Negative Differential Photoconductance in Gold Nanoparticle Arrays in the Coulomb Blockade Regime. *ACS Nano* **2012,** *6*, 4181-4189.
18. Zhang, T.; Guerin, D.; Alibart, F.; Vuillaume, D.; Lmimouni, K.; Lenfant, S.; Yassin, A.; Ocafrain, M.; Blanchard, P.; Roncali, J., Negative Differential Resistance, Memory, and Reconfigurable Logic Functions Based on Monolayer Devices Derived from Gold Nanoparticles Functionalized with Electropolymerizable TEDOT Units. *Journal of Physical Chemistry C* **2017,** *121* (18), 10131-10139.
19. Bose, S. K.; Lawrence, C. P.; Liu, Z.; Makarenko, K. S.; van Damme, R. M. J.; Broersma, H. J.; van der Wiel, W. G., Evolution of a designless nanoparticle network into reconfigurable Boolean logic. *Nature Nanotech* **2015,** *10*, 1048-1053.
20. Viero, Y.; Guerin, D.; Alibart, F.; Lenfant, S.; Vuillaume, D., Light-stimulable molecules/nanoparticles multi-terminal networks for switchable logical function and reservoir computing. *Adv. Func. Mater.* **2018,** *28*, 1801506.
21. Wang, T.; Nijhuis, C. A., Molecular electronic plasmonics. *Applied Materials Today* **2016,** *3*, 73-86.

22. Hsiao, V. K. S.; Zheng, Y. B.; Juluri, B. K.; Huang, T. J., Light-Driven Plasmonic Switches Based on Au Nanodisk Arrays and Photoresponsive Liquid Crystals. *Adv. Mater.* **2008,** *20* (18), 3528-3532.
23. Leroux, Y.; Lacroix, J.-C.; Fave, C.; Trippé, G.; Felidj, N.; Aubard, J.; Hohenau, A.; Krenn, J. R., Tunable Electrochemical Switch of the Optical Properties of Metallic Nanoparticles. *ACS Nano* **2008,** *2* (4), 728-732.
24. Zheng, Y. B.; Yang, Y.-W.; Jensen, L.; Fang, L.; Juluri, B. K.; Flood, A. H.; Weiss, P. S.; Stoddart, J. F.; Huang, T. J., Active molecular plasmonics: controlling plasmon resonances with molecular switches. *Nano Letters* **2009,** *9* (2), 819-25.
25. Stockhausen, V.; Martin, P.; Ghilane, J.; Leroux, Y.; Randriamahazaka, H.; Grand, J.; Felidj, N.; Lacroix, J.-C., Giant Plasmon Resonance Shift Using Poly(3,4-ethylenedioxythiophene) Electrochemical Switching. *J Am Chem Soc* **2010,** *132* (30), 10224-10226.
26. Zheng, Y. B.; Kiraly, B.; Cheunkar, S.; Huang, T. J.; Weiss, P. S., Incident-Angle-Modulated Molecular Plasmonic Switches: A Case of Weak Exciton–Plasmon Coupling. *Nano Lett* **2011,** *11* (5), 2061-2065.
27. Müller, M.; Jung, U.; Gusak, V.; Ulrich, S.; Holz, M.; Herges, R.; Langhammer, C.; Magnussen, O., Localized Surface Plasmon Resonance Investigations of Photoswitching in Azobenzene-Functionalized Self-Assembled Monolayers on Au. *Langmuir* **2013,** *29* (34), 10693-10699.
28. Joshi, G. K.; Blodgett, K. N.; Muhoberac, B. B.; Johnson, M. A.; Smith, K. A.; Sardar, R., Ultrasensitive Photoreversible Molecular Sensors of Azobenzene-Functionalized Plasmonic Nanoantennas. *Nano Lett* **2014,** *14*, 532-540.
29. Schaming, D.; Nguyen, V.-Q.; Martin, P.; Lacroix, J.-C., Tunable Plasmon Resonance of Gold Nanoparticles Functionalized by Electroactive Bisthienylbenzene Oligomers or Polythiophene. *J. Phys. Chem. C* **2014,** *118* (43), 25158-25166.
30. Song, H.; Jing, C.; Ma, W.; Xie, T.; Long, Y.-T., Reversible photoisomerization of azobenzene molecules on a single gold nanoparticle surface. *Chem. Commun.* **2016,** *52* (14), 2984-2987.
31. Du, W.; Wang, T.; Chu, H.-S.; Nijhuis, C. A., Highly efficient on-chip direct electronic–plasmonic transducers. *Nature Photon* **2017,** *11* (10), 623-627.
32. Ward, D. R.; Hüser, F.; Pauly, F.; Cuevas, J.-C.; Natelson, D., Optical rectification and field enhancement in a plasmonic nanogap. *Nature Nanotech* **2010,** *5*, 732-736.
33. Banerjee, P.; Conklin, D.; Nanayakkara, S.; Park, T.-H.; Therien, M. J.; Bonnell, D. A., Plasmon-induced electrical conduction in molecular devices. *ACS Nano* **2010,** *4* (2), 1019-1025.
34. Noy, G.; Ophir, A.; Selzer, Y., Response of Molecular Junctions to Surface Plasmon Polaritons. *Angew. Chem. Int. Ed.* **2010,** *49* (33), 5734-5736.
35. Arielly, R.; Ofarim, A.; Noy, G.; Selzer, Y., Accurate Determination of Plasmonic Fields in Molecular Junctions by Current Rectification at Optical Frequencies. *Nano Lett* **2011,** *11* (7), 2968-2972.
36. Zelinskyy, Y.; May, V., Photoinduced Switching of the Current through a Single Molecule: Effects of Surface Plasmon Excitations of the Leads. *Nano Lett* **2012,** *12*, 446-452.
37. Vadai, M.; Nachman, N.; Ben-Zion, M.; Bürkle, M.; Pauly, F.; Cuevas, J.-C.; Selzer, Y., Plasmon-Induced Conductance Enhancement in Single-Molecule Junctions. *J. Phys. Chem. Lett.* **2013,** *4* (17), 2811-2816.

38. Duché, D.; Palanchoke, U.; Terracciano, L.; Dang, F.-X.; Patrone, L.; Le Rouzo, J.; Balaban, T. S.; Alfonso, C.; Charai, A.; Margeat, O.; Ackermann, J.; Gourgon, C.; Simon, J.-J.; Escoubas, L., Molecular diodes in optical rectennas. *SPIE Nanoscience + Engineering* **2016,** *9929*, 99290T.
39. Tien, P. K.; Gordon, J. P., Multiphoton Process Observed in the Interaction of Microwave Fields with the Tunneling between Superconductor Films. *Phys. Rev.* **1963,** *129* (2), 647-651.
40. Tame, M. S.; McEnery, K. R.; Ozdemir, S. K.; Lee, J.; Maier, S. A.; Kim, M. S., Quantum plasmonics. *Nat Phys* **2013,** *9* (6), 329-340.
41. Zhu, W.; Esteban, R.; Borisov, A. G.; Baumberg, J. J.; Nordlander, P.; Lezec, H. J.; Aizpurua, J.; Crozier, K. B., Quantum mechanical effects in plasmonic structures with subnanometre gaps. *Nature Communications* **2016,** *7*, 11495.
42. Orgiu, E.; George, J.; Hutchison, J. A.; Devaux, E.; Dayen, J. F.; Doudin, B.; Stellacci, F.; Genet, C.; Schachenmayer, J.; Genes, C.; Pupillo, G.; Samori, P.; Ebbesen, T. W., Conductivity in organic semiconductors hybridized with the vacuum field. *Nat Mater* **2015,** *14* (11), 1123-1129.
43. Tan, S. F.; Wu, L.; Yang, J. K. W.; Bai, P.; Bosman, M.; Nijhuis, C. A., Quantum Plasmon Resonances Controlled by Molecular Tunnel Junctions. *Science (New York, NY)* **2014,** *343* (6178), 1496-1499.
44. Koya, A. N.; Lin, J., Charge transfer plasmons: Recent theoretical and experimental developments. *Appl. Phys. Rev.* **2017,** *4* (2), 021104.
45. Mahmoud, M. A., Electromagnetic Field of Plasmonic Nanoparticles Extends the Photoisomerization Lifetime of Azobenzene. *J. Phys. Chem. C* **2017,** *121* (33), 18144-18152.
46. Tsuboi, Y.; Shimizu, R.; Shoji, T.; Kitamura, N., Near-Infrared Continuous-Wave Light Driving a Two-Photon Photochromic Reaction with the Assistance of Localized Surface Plasmon. *J Am Chem Soc* **2009,** *131* (35), 12623-12627.
47. Garcia-Amorós, J.; Swaminathan, S.; Sortino, S.; Raymo, F. M., Plasmonic Activation of a Fluorescent Carbazole–Oxazine Switch. *Chem. Eur. J.* **2014,** *20* (33), 10276-10284.
48. Impellizzeri, S.; Simoncelli, S.; Hodgson, G. K.; Lanterna, A. E.; McTiernan, C. D.; Raymo, F. M.; Aramendia, P. F.; Scaiano, J. C., Two-Photon Excitation of a Plasmonic Nanoswitch Monitored by Single-Molecule Fluorescence Microscopy. *Chem. Eur. J.* **2016,** *22* (21), 7281-7287.
49. Karpe, S.; Ocafrain, M.; Smaali, K.; Lenfant, S.; Vuillaume, D.; Blanchard, P.; Roncali, J., Oligothiophene-derivatized azobenzene as immobilized photoswitchable conjugated systems. *Chemical Communications* **2010,** *46* (21), 3657-3659.
50. Smaali, K.; Lenfant, S.; Karpe, S.; Oçafrain, M.; Blanchard, P.; Deresmes, D.; Godey, S.; Rochefort, A.; Roncali, J.; Vuillaume, D., High On−Off Conductance Switching Ratio in Optically-Driven Self-Assembled Conjugated Molecular Systems. *ACS Nano* **2010,** *4* (4), 2411-2421.
51. Johansson, P.; Xu, H.; Käll, M., Surface-enhanced Raman scattering and fluorescence near metal nanoparticles. *Phys. Rev. B* **2005,** *72* (3), 035427.
52. Muhlschlegel, P.; Eisler, H. J.; Martin, O. J. F.; Hecht, B.; Pohl, D. W., Resonant Optical Antennas. *Science (New York, NY)* **2005,** *308* (5728), 1607.
53. Feringa, B. L., *Molecular Switches*. Wiley-VCH Verlag GmbH: 2001.

54. Mangold, M. A.; Calame, M.; Mayor, M.; Holleitner, A. W., Resonant Photoconductance of Molecular Junctions Formed in Gold Nanoparticle Arrays. *J Am Chem Soc* **2011,** *133* (31), 12185-12191.
55. Zhou, J.; Wang, K.; Xu, B.; Dubi, Y., Photoconductance from Exciton Binding in Molecular Junctions. *J Am Chem Soc* **2017,** *140* (1), 70-73.
56. Alemani, M.; Peters, M. V.; Hecht, S.; Rieder, K.-H.; Moresco, F.; Grill, L., Electric Field-Induced Isomerization of Azobenzene by STM. *J Am Chem Soc* **2006,** *128* (45), 14446-14447.
57. Li, J.; Cushing, S. K.; Meng, F.; Senty, T. R.; Bristow, A. D.; Wu, N., Plasmon-induced resonance energy transfer for solar energy conversion. *Nature Photon* **2015,** *9*, 601-608.
58. Gholamrezaie, F.; Vijayaraghavan, R. K.; Meskers, S. C. J., Photovoltaic action in a self-assembled monolayer of hemicyanine dyes on gold from dissociation of surface plasmons. *Appl. Phys. Lett.* **2015,** *106* (18), 183303.
59. Hush, N. S.; Wong, A. T.; Bacskay, G. B.; Reimers, J. R., Electron and energy transfer through bridged systems. 6. Molecular switches: the critical field in electric field activated bistable molecules. *J Am Chem Soc* **1990,** *112* (11), 4192-4197.
60. Klajn, R.; Wesson, P. J.; Bishop, K. J. M.; Grzybowski, B. A., Writing self-erasing images using metastable nanoparticle "inks". *Angew. Chem. Int. Ed.* **2009,** *48* (38), 7035-7039.
61. Bourgoin, J.-P.; Kergueris, C.; Lefèvre, E.; Palacin, S., Langmuir–Blodgett films of thiol-capped gold nanoclusters: fabrication and electrical properties. *Thin Solid Films* **1998,** *327-329*, 515-519.
62. Santhanam, V.; Liu, J.; Agarwal, R.; Andres, R. P., Self-Assembly of Uniform Monolayer Arrays of Nanoparticles. *Langmuir* **2003,** *19*, 7881-7887.
63. Monti, S.; Orlandi, G.; Palmieri, P., Features of the photochemically active state surfaces of azobenzene. *Chemical Physics* **1982,** *71* (1), 87-99.

Supporting information

# Electrical detection of plasmon-induced isomerization in molecule-nanoparticle network devices

Didier Stiévenard,[1] David Guérin,[1] Stéphane Lenfant,[1] Gaëtan Lévêque,[1] Christian A. Nijhuis[2,3,4] & DominiqueVuillaume[1]*.

1) Institut d'Electronique, Microélectronique et Nanotechnologie (IEMN), CNRS, Université de Lille,
Avenue Poincaré, F-59652cedex, Villeneuve d'Ascq, France.

2) Department of Chemistry, National University of Singapore, 3 Science Drive 3, 117543 Singapore, Singapore.

3) Centre for Advanced 2D Materials and Graphene Research Centre, National University of Singapore, 6 Science Drive 2, 117546 Singapore, Singapore.

4) NUSNNI-Nanocore, National University of Singapore, Singapore 117411, Singapore.

* corresponding author : dominique.vuillaume@iemn.fr

## 1. 2D-NPSAN : corrected current

For 2D NPSANs, it was not possible to obtain a stable current during the pre-configuration, the dark period and the LSP excitation. Figure S1 shows a typical current vs. time recorded for a sequence : pre-configuration in *trans* (470 nm light), dark, LSP excitation (590 nm) and dark. To extract the corrected current during PII shown in Fig. 3-a (main text), we subtracted a liner fit (blue line) from the raw data. Figure S2 shows the data when the AzBT are pre-configured in their *cis* state, and the fitted base line used to calculate the corrected current shown in Fig. 3-b.

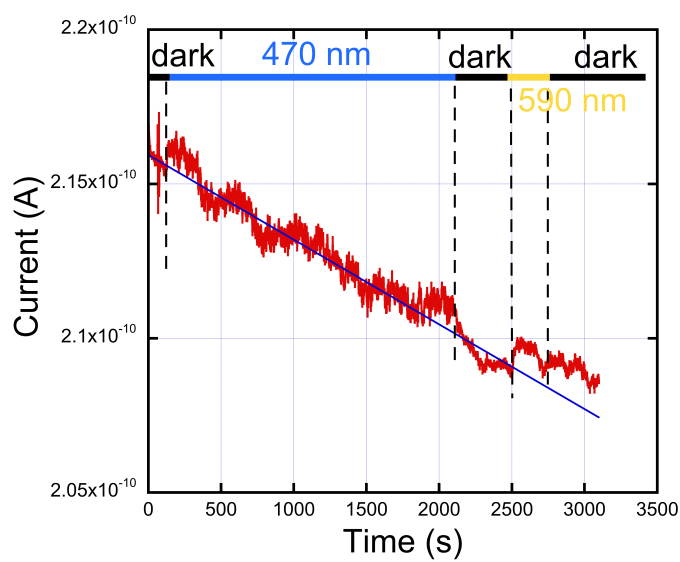

*Figure S1.* *Measured current for the 2D NSPSAN under a sequence : pre-configuration in trans (470 nm light), dark, LSP excitation (590 nm) and dark. The blue line is a linear function : $i = 2.16 \times 10^{-10} - 2.74 \times 10^{-15} \, t$, fitted for $0 < t < 2500$ and extrapolated to subtract the base line in Fig. 3-a.*

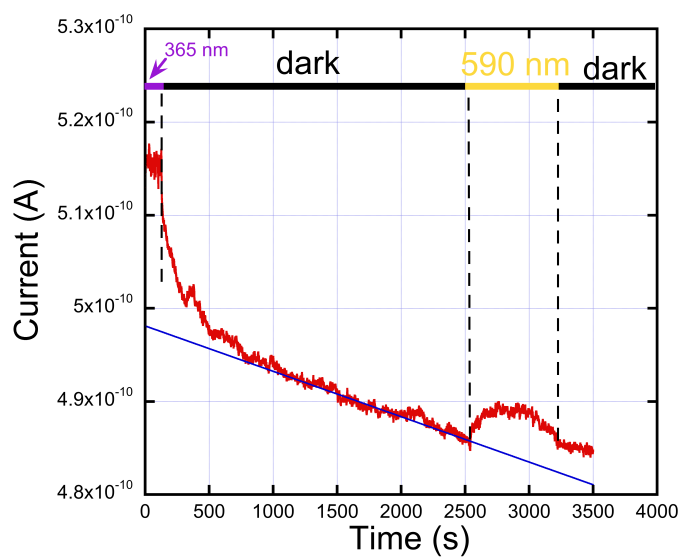

*Figure S2.* *Measured current for the 2D NSPSAN under a sequence : pre-configuration in cis (365 nm light), dark, LSP excitation (590 nm) and dark. The blue line is a linear function : $i = 4.98 \times 10^{-10} - 4.87 \times 10^{-15} \, t$, fitted for $1000 < t < 2500$ and extrapolated to subtract the base line in Fig. 3-b.*

## 2. UV-vis isomerization.

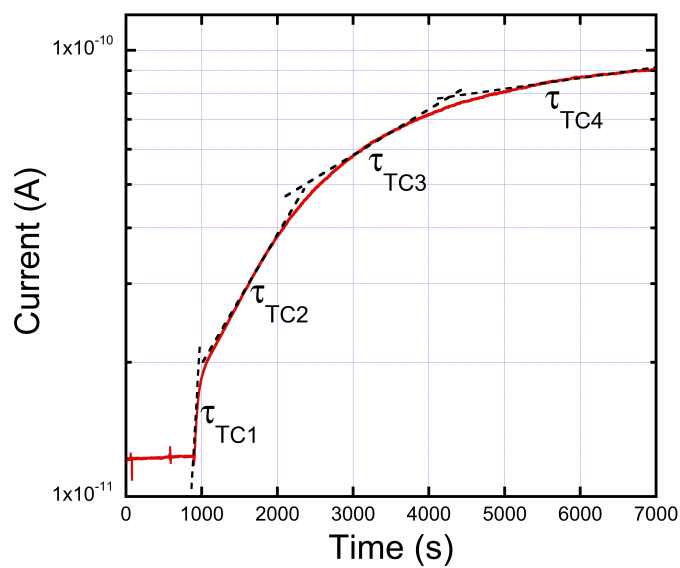

***Figure S3.*** *Semilog plot of the trans to cis isomerization for the 2D-NPSANs (data form Fig. 4-a). We can extract several time constants : $\tau_{TC1}$ = 100 s, $\tau_{TC2}$ = 1360, $\tau_{TC3}$ = 1650 and $\tau_{TC4}$ = 2.2x10$^4$ s*

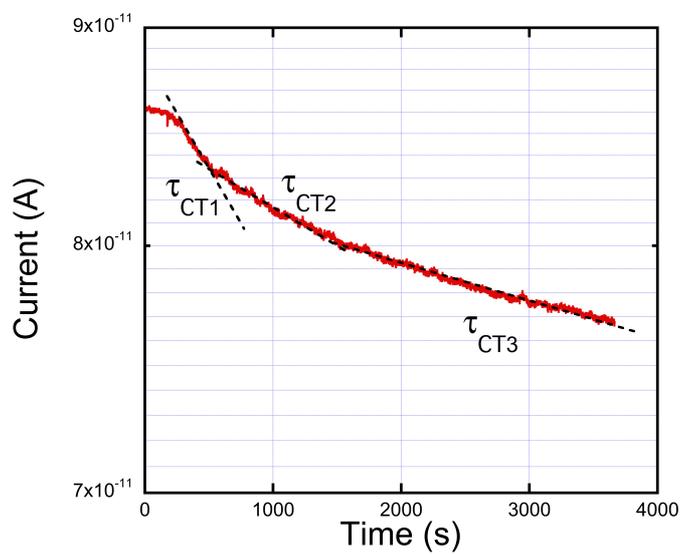

***Figure S4.*** *Semilog plot of the cis to trans isomerization for the 2D-NPSANs (data form Fig. 4-b). We can extract several time constants : $\tau_{CT1}$ = 8900 s, $\tau_{CT2}$ = 3x10$^4$ s and $\tau_{CT3}$ = 5.4x10$^4$ s.*

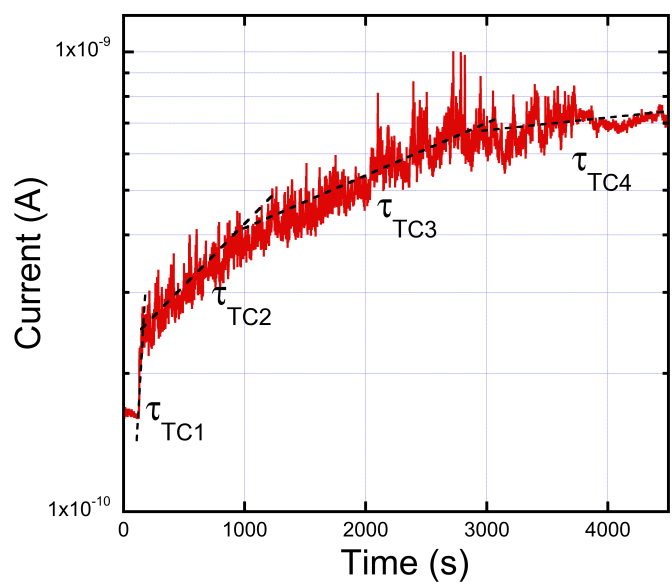

*Figure S5.* Semilog plot of the trans to cis isomerization for the cluster-NPSANs (data form Fig. 4-c). We can extract several time constants : $\tau_{TC1}$ = 30 s, $\tau_{TC2}$ = 1570 $\tau_{TC3}$ = 2900 and $\tau_{TC4}$ = 1.35x10$^4$ s.

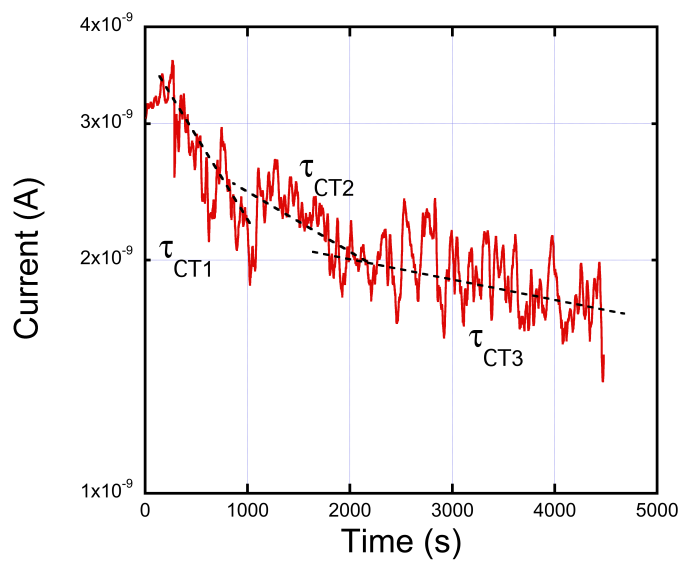

*Figure S6.* Semilog plot of the cis to trans isomerization for the cluster-NPSANs (data form Fig. 4-d). We can extract several time constants : $\tau_{CT1}$ = 1950 s, $\tau_{CT2}$ = 4200 and $\tau_{CT3}$ = 1.2x10$^4$ s.

For the 2D-NPSANs, the *trans* to *cis* isomerization reveals several features, starting by a "fast" kinetics with a time constant of the order of $\tau_{TC1}$ = 100 s, followed by slower kinetics characterized by multi-exponential behaviors with time constants larger than ~$10^3$ s (Fig. S3). For the *cis* to *trans* isomerisation (Fig. S4) three typical time constants are extracted, with a first time constant $\tau_{CT1}$ = 8900 s much slower than for the *trans* to *cis* case. The same behavior is observed for the cluster-NSANs (Figs. S5 and S6). In the cluster-NPSANs the current levels are higher (as expected due to a more important number of current pathways), again with several time constants. The extracted time constants for the *trans*-to-*cis* isomerisation are of the order of $\tau_{TC1}$ = 30 s (Fig. S5), again followed by several time constants > $10^3$ s. For the *cis* to *trans* isomerization, we have $\tau_{CT1}$ = 1950 s (Figure S6) for the first time constant. We note that these multi-exponential behaviors of the isomerization kinetics are quite similar for the 2D- and cluster-NPSANs, suggesting that the NP organizations are similar in the two types of devices, or that NP organization is of little importance to influence the AzBT isomerization. Table 1 summarizes all the measured time constants. We note that the multi-exponential behaviors of the UV-vis isomerization of AzBT in NPSANs differ from the previously observed behavior (single time constant) of the same molecules in a self-assembled monolayer, albeit an initial faster kinetics with a small amplitude is also observed for AzBT in SAMs.[1] This feature is likely related to a more disordered molecular structure in NPSANs.

| | 590 nm | | 365 nm trans to cis | | 470 nm cis to trans |
|---|---|---|---|---|---|
| | trans state (s) | cis state (s) | $\tau_{TC1}$ (s) | $\tau_{TCi}$ (s), i≥2 | $\tau_{CTi}$ (s), i≥1 |
| 2D-NPSAN | 22/63 | n.a. | 100 | 1360, 1650, $2.2 \times 10^4$ | 8900, $3 \times 10^4$, $5.4 \times 10^4$ |
| cluster-NPSAN | 90 | 103 | 30 | 1570, 2900, $1.35 \times 10^4$ | 1950, 4200, $1.2 \times 10^4$ |

**Table 1**. Summary of the typical measured time constants under the 590 nm light irradiation for the cluster-NPSANs with the AzBT molecules in the *trans* state and *cis* state (Figs. 2-a and 2-b); similar data from Figs. 3-a and 3-b for the 2D-NPSANs with the AzBT molecules in the *trans* state only (see main text); and for both the 2D-NPSANs and cluster-NPSANs (data from Figs. 4 and S3-S6) under a 365 nm light irradiation (*trans* to *cis* isomerisation) and 470 nm light irradiation (*cis* to *trans* isomerisation).

## 3. PAT current.

The PAT current is calculated from the amplification factor M of the local electric field induced by the LSP between NPs for various NP sizes (10-60 nm) and gap distance between adjacent NPs (1-5 nm). These simulations show (next section 4, Fig. S8) that M does not overcome a value of ~10 for NPs with a diameter of 10 nm, spaced by an average gap of 4.5 nm and up to ca. 80 for a gap of around 1 nm. With a light power of 11.5 mW at 590 nm, the incident electric field $E_0$ is ~ $2.9 \times 10^{-2}$ V/cm, as calculated from $P = 0.5(\varepsilon/\mu)^{0.5}E^2$ with P the light power, $\varepsilon$ and $\mu$ the permittivity and permeability in vacuum. Thus, we estimated $E_\omega = ME_0 = 2.9 \times 10^{-1}$ V/cm. Therefore, as the nano-gap between two NP is ~ 4.5 nm,[2] the average LSP electric field $V_\omega$ across the NP-molecule-NP gap is of the order of $1.3 \times 10^{-7}$ V. We calculated an upper limit of the PAT-induced change in the molecular conductance $\Delta G_{mol} = G_{dc}(\omega) - G_{dc}(0) = 8.25 \times 10^{-20}$ S using Eq. (1) in the main text, considering that the transmission coefficient though the molecule τ(EF

± ℏω) is 1 at the maximum. We converted $\Delta G_{mol}$ to $\Delta I$, the PAT-induced current in the NPSAN, according to the following approach. Taking into account that about 200 molecules are inserted in the nanogap between two adjacent NPs (see Molecular Dynamics calculations in Ref. 2), we have $\Delta G_J = 200\Delta G_{mol}$ with $\Delta G_J$ the conductance of one NP-molecules-NP junction. Assuming a hexagonal compact organization of the NPs in the network, the sheet conductance in the NPSAN is $\Delta G_\square = (2/\sqrt{3})\Delta G_J$, with $\Delta G_\square = \Delta G * L/W$; with the NPSAN conductance $\Delta G = \Delta I/V$ (V the dc applied voltage, 2.5 V), L (150 nm) the length between the electrodes and W (40 nm) the width (see Fig. 1-c). Thus, from $\Delta G_{mol} = 8.25 \times 10^{-20}$ S, we estimated an upper limit for the PAT-induced current $\Delta I = 1.3 \times 10^{-17}$ A.

## 4. UV-vis spectrum

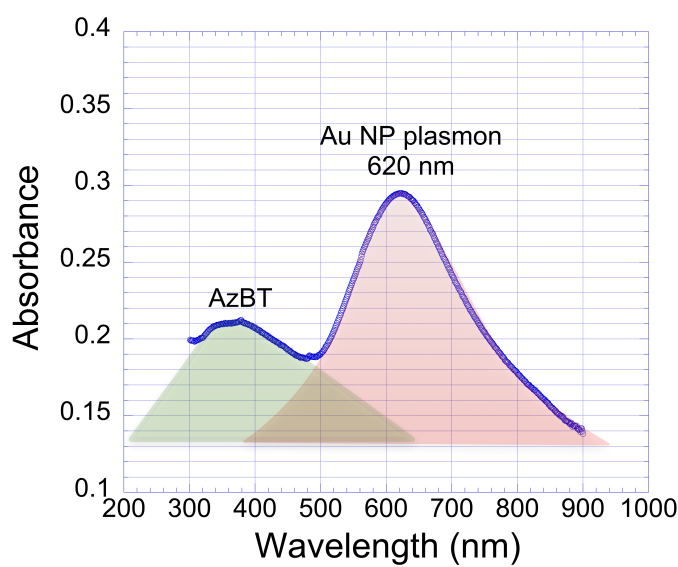

*Figure S7.* *UV-vis absorbance measurement for the trans-AzBT NPSAN deposited on quartz following the same methods as for NSPANs deposited on SiO$_2$ substrate used for electrical measurements (see Methods in the main paper). A transparent*

*substrate is required for this measurement. Green and red areas are guides for eyes to highlight the contribution of the Au nanoparticules (LSP) and the AzBT (centered at 350-360 nm, see absorption of the AzBT molecules in Refs. 1, 3).*

## 5. LSP simulations

Optical numerical simulations have been performed with a commercial finite element software (Comsol) on a modeled cluster-NPSAN. It consists in gold nanoparticles with 10nm diameter packed in a three-layer closed-packed structure of FCC type, with interparticle separation fixed to 4.5 nm. The cluster-NPSAN lies on a silica substrate, and is illuminated by a plane wave in oblique incidence (45°) from air. Transmission is computed in silica. The dielectric constant of gold is from Johnson and Christy,[4] and size correction has been applied as the diameter of the nanospheres is comparable to the mean free path of the conduction electron in gold.[5] The inter-particle medium is modeled as homogeneous and isotropic, and its dielectric constant has been adjusted in order to match the resonance wavelength of 630 nm obtained in UV-vis measurements (Fig. S7), which led to a value of 3.8, in agreement with literature for organic materials based on azobenzene derivatives.[6] The obtained M-factor is about 6.5 at resonance.

Then, we have estimated numerically the M-factor as a function of the gap and radius in a dimer of gold nanospheres embedded in an homogeneous medium of dielectric constant 3.8, at a fixed wavelength of 590 nm corresponding to the experimental LSP excitation. Simulations were done using a multi-particle Mie method.[7] The enhancement factor M reaches about 80 for a gap of 1 nm and radius of 15 nm, and drops to 11 for a gap of 4.5 nm and radius 5nm in our case (red cross). This value is slightly larger than in the modeled NPSAN due to increased confinement of light within the dimer.

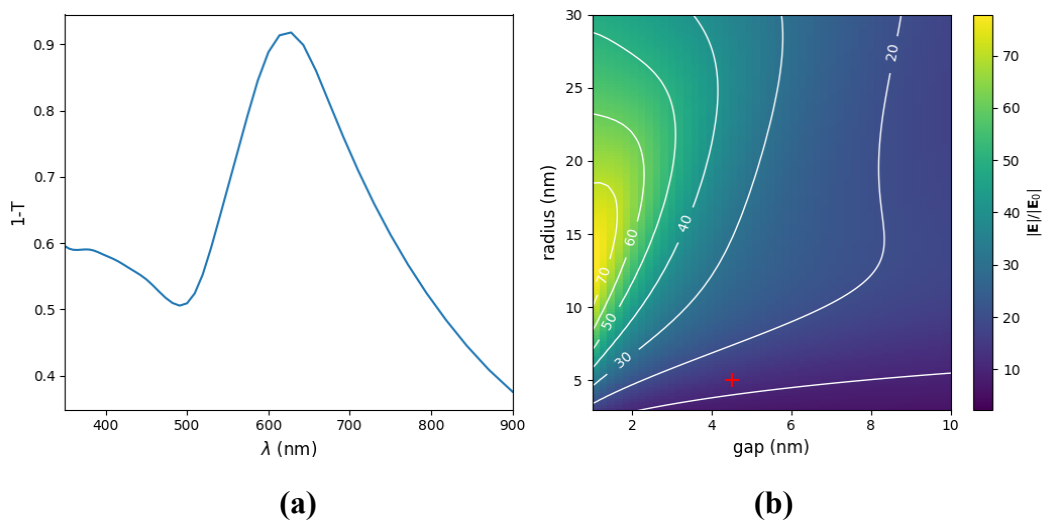

(a)                           (b)

*Figure S8: (a) Extinction spectrum of the modeled three-layer cluster-NSPAN, with a dielectric constant of 3.8 for the interparticle medium. (b) M-factor as a function of gap and radius of two gold nanospheres in a homogeneous medium with refractive index 3.8. The field is computed 0.2nm away from one nanosphere surface, inside the gap. The red cross marks a radius of 5 nm and gap 4.5 nm.*

## 6. Electrical switching of AzBT

In STM studies reported in Ref. 8, the molecules are lying flat on Au(111) surface, and the samples are in UHV and at 5K. Such configuration and conditions are quite different from our case. Moreover, during the STM measurements, the electric field is applied perpendicular to the surface/molecule whereas in our case, various angles exist between the plasmonic electric field and the molecules covering the Au NPs. This is of prime importance because the orientation of the molecule (i.e. its permanent dipole and the component of its polarizability tensor along the electric field direction) is a key factor for the isomerization.[9] We also note that for Au(100) or Cu(111) surfaces,[10] no switching effect was detected, showing that these observations are surface dependent, and, to the best of our knowledge, electrical switching of azobenzene has only been reported in Ref. 8. Finally, STM experiments used a static electric field compared to a dynamic one in our case.

## 7. AFM images.

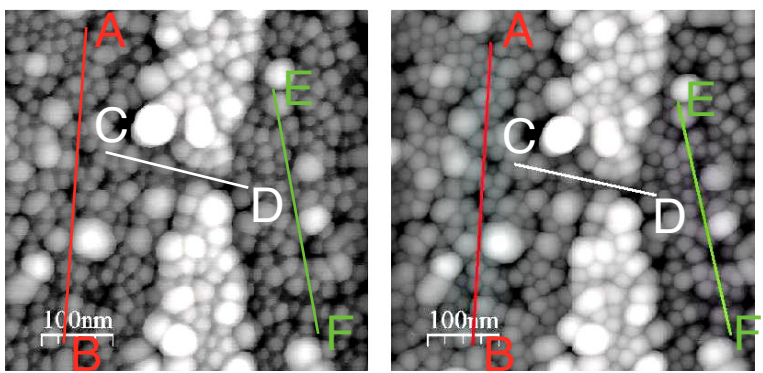

*Figure S9.* AFM images of the NPSANs with the AzBT in the trans isomer (left) and cis isomer (right). We do not observed a significant modification of the NP network morphology. We only note some random distance variations, which may be due fluctuations during the long time of experiments (light illumination and AFM images). Formation example, the point-to-point distances AB (421 nm), CD (305 nm) and EF (337 nm) vary by -4.9%, -0.1% and +4.1%, between the trans and cis isomers, respectively. This ±5% relative variation applied to the inter-NP distance correspond to ±0.22 nm, i.e below the statistical size distribution of ± 1 nm.[2]

## References

1.     Smaali, K.; Lenfant, S.; Karpe, S.; Oçafrain, M.; Blanchard, P.; Deresmes, D.; Godey, S.; Rochefort, A.; Roncali, J.; Vuillaume, D. *ACS Nano* **2010,** 4, (4), 2411-2421.
2.     Viero, Y.; Copie, G.; Guerin, D.; Krzeminski, C.; Vuillaume, D.; Lenfant, S.; Cleri, F. *J. Phys. Chem. C* **2015,** 119, 21173-21183.

3. Karpe, S.; Ocafrain, M.; Smaali, K.; Lenfant, S.; Vuillaume, D.; Blanchard, P.; Roncali, J. *Chemical Communications* **2010,** 46, (21), 3657-3659.
4. Johnson, P. B.; Christy, R. W. *Phys. rev. B* **1972,** 6, 4370.
5. Berciaud, S.; Cognet, L.; Tamarat, P.; Lounis, B. *Nano Lett* **2005,** 5, 515-518.
6. Stangenberg, R.; Grigoriadis, C.; Butt, H.-J.; Müllen, K.; Floudas, G. *Colloid and Polymer Science* **2014,** 292, (8), 1939-1948.
7. Xu, Y.-l.; Gustafson, B. Journal of Quantitative Spectroscopy & Radiative Transfer **2001,** 70, 395-419.
8. Alemani, M.; Peters, M. V.; Hecht, S.; Rieder, K.-H.; Moresco, F.; Grill, L. *J Am Chem Soc* **2006,** 128, (45), 14446-14447.
9. Füchsel, G.; Klamroth, T.; Dokić, J.; Saalfrank, P. *J Phys Chem B* **2006,** 110, (33), 16337-16345.
10. Alemani, M.; Selvanathan, S.; Ample, F.; Peters, M. V.; Rieder, K.-H.; Moresco, F.; Joachim, C.; Hecht, S.; Grill, L. *J. Phys. Chem. C* **2008,** 112, (28), 10509-10514.